\documentclass{icrc2009}

\usepackage{graphicx}
\usepackage[caption=false]{caption}
\usepackage[font=footnotesize]{subfig}
\usepackage{fixltx2e}
\usepackage{url}

\newcommand{\shorttitle}[1]%
{\markboth{Proceedings of the 31\MakeLowercase{$^{st}$} ICRC, {\L}\'{o}d\'{z} 2009}{#1} }
\newcommand{\etal}{\MakeLowercase{\textit{et al. }}} 

\begin{document}
\title{A first joint M87 campaign in 2008 from radio to TeV gamma-rays}

\author{\IEEEauthorblockN{R. M. Wagner\IEEEauthorrefmark{1},
			  M. Beilicke\IEEEauthorrefmark{2},
			  F. Davies\IEEEauthorrefmark{3},
			  H. Krawczynski\IEEEauthorrefmark{2},
                          D. Mazin\IEEEauthorrefmark{4},
                          M. Raue\IEEEauthorrefmark{5},
\\
                          S. Wagner\IEEEauthorrefmark{6} and
                          R. C. Walker\IEEEauthorrefmark{3}
for the H.E.S.S.\IEEEauthorrefmark{7}, MAGIC\IEEEauthorrefmark{8} and VERITAS\IEEEauthorrefmark{9} collaborations\\and the VLBA 43 GHz M87 monitoring team}
                            \\
\IEEEauthorblockA{\IEEEauthorrefmark{1}Max-Planck-Institut f\"ur Physik, D-80805 Munich, Germany}
\IEEEauthorblockA{\IEEEauthorrefmark{2}Washington University in St. Louis/McDonnell Center for the Space Sciences, St. Louis, MO 63130, USA}
\IEEEauthorblockA{\IEEEauthorrefmark{3}National Radio Astronomy Observatory, Socorro, NM 87801, USA}
\IEEEauthorblockA{\IEEEauthorrefmark{4}IFAE, E-08193 Bellaterra (Barcelona), Spain}
\IEEEauthorblockA{\IEEEauthorrefmark{5}Max-Planck-Institut f\"ur Kernphysik, D-69117 Heidelberg, Germany}
\IEEEauthorblockA{\IEEEauthorrefmark{6}Landessternwarte Heidelberg, D-69117 Heidelberg, Germany}
\IEEEauthorblockA{\IEEEauthorrefmark{7}\texttt{http://www.mpi-hd.mpg.de/hfm/HESS/pages/collaboration/institutions/}}
\IEEEauthorblockA{\IEEEauthorrefmark{8}\texttt{http://wwwmagic.mppmu.mpg.de/collaboration/members/}}
\IEEEauthorblockA{\IEEEauthorrefmark{9}\texttt{http://veritas.sao.arizona.edu/conferences/authors?icrc2009}}
}

\shorttitle{Wagner \etal Joint M\,87 multiwavelength campaign in 2008}
\maketitle

\begin{abstract}
M87, the central galaxy of the Virgo cluster, is the first radio galaxy
detected in the TeV regime.  The structure of its jet, which is not pointing
toward the line of sight, is spatially resolved in X-ray (by Chandra), in
optical and in radio observations.  Time correlation between the TeV flux and
emission at other wavelengths provides a unique opportunity to localize the
very high energy gamma-ray emission process occurring in AGN. For 10 years, M87
has been monitored in the TeV band by atmospheric Cherenkov telescopes. In
2008, the three main atmospheric Cherenkov telescope observatories (H.E.S.S.,
MAGIC and VERITAS) coordinated their observations in a joint campaign from
January to May with a total observation time of $\approx$~120 hours.  The
campaign largely overlapped with an intensive VLBA project monitoring the core
of M87 at 43\,GHz every 5 days.  In February, high TeV activities with rapid
flares have been detected. Contemporaneously, M87 was observed with high
spatial resolution instruments in X-rays (Chandra).  We discuss the results of
the joint observation campaign in 2008.
\end{abstract}


\section{Introduction: M\,87 is a unique laboratory}

Active galactic nuclei (AGN) are extreme extragalactic objects showing
core-dominated emission (broadband continuum ranging from radio to X-ray
energies) and strong variability on different timescales. A supermassive black
hole (in the center of the AGN) surrounded by an accretion disk is believed to
power the relativistic plasma outflows (jets) which are found in many AGN.
Around 25 AGN have been found to emit VHE $\gamma$-rays (E$>$100\,GeV).  The
size of the VHE $\gamma$-ray emission region can generally be constrained by
the time scale of the observed flux variability \cite{Mrk421_Burst,Aha06} but
its location remains unknown. 

The giant radio galaxy M\,87 is located at a distance of $16 \, \mathrm{Mpc}$
(50 million light years) in the Virgo cluster of galaxies \cite{Mac99}. The
angle between the plasma jet in M\,87 and the line of sight is estimated to lie
between $20^{\circ}-40^{\circ}$ \cite{Bir95,Bir99}.  With its proximity, its
bright and well resolved jet, and its very massive black hole with $3 \times
10^{9} \, \rm{M}_{\odot}$, \cite{BH_M87}, M\,87 provides an excellent
opportunity to study the inner structures of the jet, which are expected to
scale with the gravitational radius of the black hole. Substructures of the jet
are resolved in the X-ray, optical and radio wavebands \cite{ChandraSpecM87}
and high-frequency radio very long baseline interferometry (VLBI) observations
with sub-milliarcsecond (mas) resolution are starting to probe the collimation
region of the jet \cite{JetFormation}. This makes M\,87 a unique laboratory in
which to study relativistic jet physics in connection with the mechanisms of
VHE $\gamma$-ray emission in AGN.

VLBI observations of the M\,87 inner jet show a well resolved, edge-brightened
structure extending to within $0.5 \, \rm{mas}$ ($0.04 \, \rm{pc}$ or $140$
Schwarzschild radii $R_{\rm{s}}$) of the core. Closer to the core, the jet has
a wide opening angle suggesting that this is the collimation region
\cite{JetFormation}.  Along the jet, monitoring observations show both
near-stationary components \cite{InnerJet} (pc-scale)  and features that move
at apparent superluminal speeds \cite{HST_Superluminal,Jet_And_TeV} (100
pc-scale). The presence of superluminal motions and the strong asymmetry of the
jet brightness indicate that the jet flow is relativistic. The near-stationary
components could be related to shocks or instabilities that can be either
stationary or move more slowly than the bulk flow.

\section{Ten years of VHE $\gamma$-ray observations of M\,87}

Currently, there are about 25 extragalactic objects~-- all belonging to the
class of AGN~-- that have been established as VHE $\gamma$-ray emitters by
ground-based imaging atmospheric Cherenkov telescopes (IACTs), such as
H.E.S.S., MAGIC and VERITAS.  So far, all of them except the radio galaxies
M\,87, Centaurus A \cite{CenA}, and possibly 3C~66B \cite{3C66B} belong to the
class of blazars (exhibiting a plasma jet pointing closely to our line of
sight). 

\begin{figure}

\includegraphics[width=0.48\textwidth]{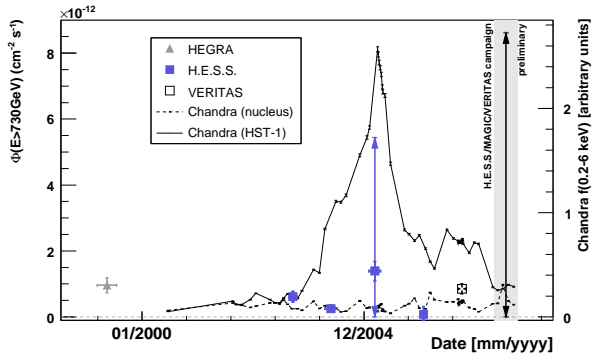}

\caption{The yearly averaged VHE $\gamma$-ray light curve $\Phi(E>730 \,
\rm{GeV})$ of M\,87, covering the 10-year period from 1998-2008. The data
points are taken from \cite{Aha03,Aha06,Acc08b,Alb08}. Vertical arrows indicate
the measured flux ranges for the given period (if variable emission was found).
Strong variabilities ($<$ 2 days flux doubling times) were measured in
$\gamma$-rays in 2005 and 2008.  The Chandra X-ray light curves of the nucleus
and HST-1 are also shown \cite{Har06}.}

\label{fig:LC_All}

\end{figure}

A first indication of VHE $\gamma$-ray emission ($> 4 \, \sigma$) from the
direction of M\,87 in 1998/99 was reported by HEGRA \cite{Aha03}. The VHE
$\gamma$-ray emission was confirmed by H.E.S.S. \cite{Aha06}, establishing
M\,87 as the first non-blazar extragalactic VHE $\gamma$-ray source. The
reported day-scale variability strongly constrains the size of the $\gamma$-ray
emission region.  VERITAS detected M\,87 in 2007 \cite{Acc08b} but with no
variability.  Recently, the short-term variability in M87 was confirmed by
MAGIC in a strong VHE $\gamma$-ray outburst \cite{Alb08,Err09}. The yearly
averaged VHE $\gamma$-ray light curve of M\,87 for the past 10 years is shown
in Fig.~\ref{fig:LC_All}.  The measured flux variability rules out large-scale
emission from dark matter annihilation \cite{Bal99}, or cosmic-ray interactions
\cite{Pfr03}. Leptonic \cite{Geo05,Len08} and hadronic \cite{Rei04} jet
emission models have been proposed to explain the TeV emission.  The location
of the VHE $\gamma$-ray emission is still unknown, but the nucleus
\cite{Ner07}, the inner jet \cite{Geo05,Rei04,Len08,Tav08} or larger structures
in the jet \cite{Che07} have been proposed as possible sites. The 2005 VHE
$\gamma$-ray flare (H.E.S.S.) was detected during an exceptional, several years
lasting X-ray outburst of the innermost knot in the jet 'HST-1' \cite{Har06},
whereas the recent VHE $\gamma$-ray flaring activity (reported here) occurred
during an X-ray low state of HST-1 (see Fig.~\ref{fig:LC_All}). In this paper
we report on a joint VHE observation campaign of M\,87 performed by H.E.S.S.,
MAGIC and VERITAS in 2008.

\begin{figure}

\centering
\includegraphics[width=0.45\textwidth]{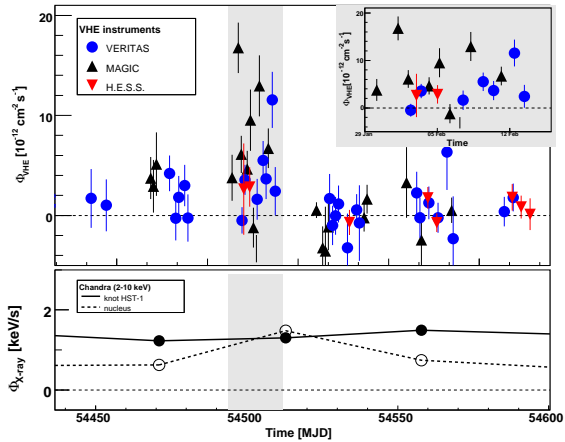}

\caption{{\it Top panel}: the night-by-night averaged VHE $\gamma$-ray light
curve $\Phi(E>350 \, \rm{GeV})$ of M\,87, covering the 2008 joint campaign.
Strong variability resulted in a detection of at least two flares by MAGIC and
VERITAS (see the zoom in part of the campaign in the inset).  The joint
H.E.S.S./MAGIC/VERITAS campaign allowed for the unprecedended coverage at VHE
$\gamma$-rays.  {\it Bottom panel}: Corresponding \textit{Chandra} measurements
of the core and the HST-1 knot of M87.
}

\label{fig:LC_2008}

\end{figure}

\section{The 2008 campaign on M87}

\subsection{Joint H.E.S.S./MAGIC/VERITAS VHE observations}

IACTs measure very high energy ($E > 100$~GeV) $\gamma$-rays.  The angular
resolution of $\sim$0.1$^\circ$ of IACTs does not allow to resolve the M\,87
jet, but the time scale of the VHE flux variability constrains the size of the
emission region, while flux correlations with observations at other wavelengths
may enable conclusions on the location of the VHE $\gamma$-ray source.  The
current generation of instruments requires less than $10 \, \rm{h}$ for the
detection of a faint source with a flux level of a few percent of the Crab
nebula flux.  For a variable VHE $\gamma$-ray source like M\,87, a joint
observation strategy as well as combining the results from several IACT
experiments (and observations at other wavelengths) can substantially improve
the scientific output (e.g.  \cite{Maz05}). Coordinated observations with the
VHE instruments H.E.S.S. \cite{Hin04}, MAGIC \cite{Lor04} and VERITAS
\cite{Acc08a} result in:  (1) an extended energy range by combining data sets
taken under different zenith angles;  (2) an extended visibility during one
night, because the visibility of any given celestial object depends on the
longitude of the experimental site;  (3) an improved overall exposure and
homogeneous coverage of the source; (4) alerts and direct follow-up
observations in case of high flux states.

\begin{figure*}
\includegraphics[width=0.45\textwidth]{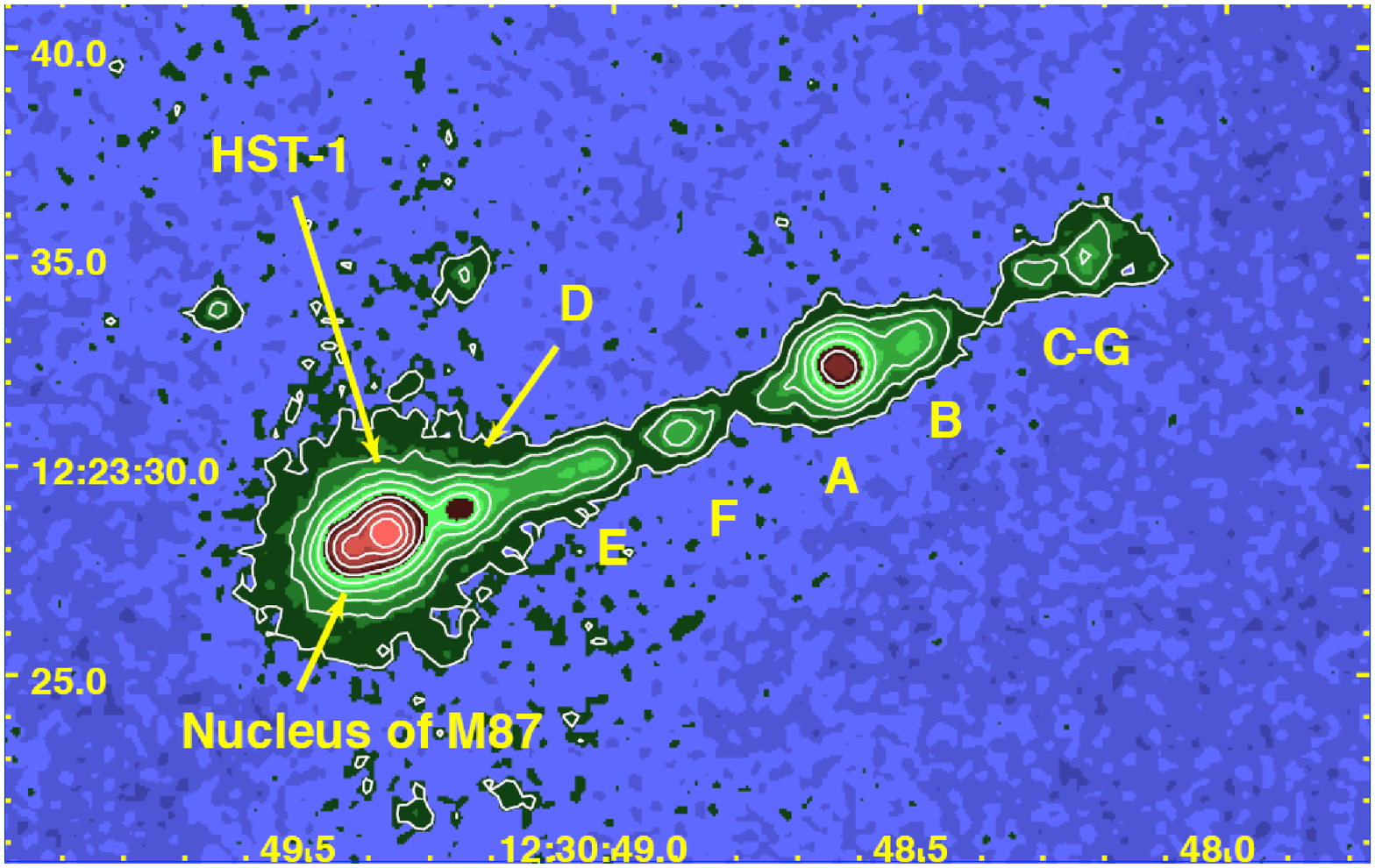}
\includegraphics[width=0.54\textwidth]{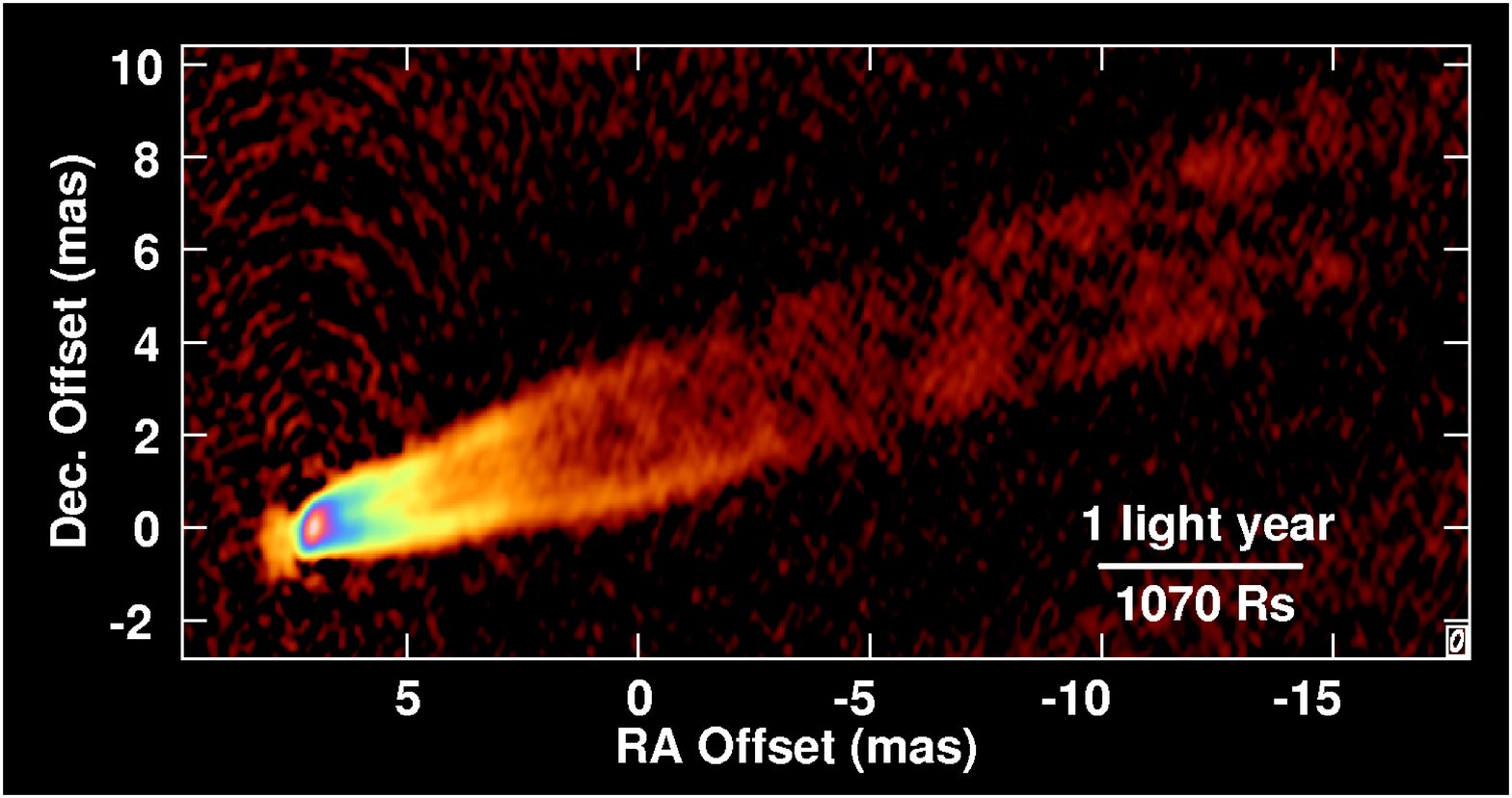}
\caption{Image of the M\,87 jet with high resolution instruments: Large-scale
jet in X-ray obtained with Chandra (left). Inner jet in radio (43\,GHz) obtained with VLBA (right).
\label{fig:X}}
\end{figure*}

Among other AGN, the radio galaxy M\,87 is part of a multi-collaboration AGN
trigger agreement between H.E.S.S., MAGIC and VERITAS. In order to achieve a
best possible VHE coverage (especially during Chandra X-ray observations) a
closer cooperation was conducted for the 2008 observations of M\,87: The
collaborations agreed to have a detailed exchange/synchronization of their
M\,87 observation schedules.  Further on, information about the status of the
observations (i.e. loss of observation time due to bad weather conditions,
etc.) was exchanged on a regular basis between the shift crews and the
observation coordinators. 

M\,87 was observed by the three experiments for a total of $>120 \, \rm{h}$ in
2008 ($\sim$95~h after quality selection). The amount of data resulted in an
unprecedentedly good coverage with $>$50 nights between January and May 2008. 

\subsection{Chandra X-ray observations}
Chandra monitoring of M\,87 began in 2002 and continues to date. The angular
resolution ($\approx$\,0.8$^{\prime\prime}$) allows resolving the large-scale
jet structure, and in particular to distinguish emission from the core and the
innermost knot `HST-1' (left panel in Fig.~\ref{fig:X}). During an observing
season, M\,87 is observed every $\sim$6~weeks. That sampling allows detection
of 1.5\,month-scale variability.  The most remarkable discovery of the
monitoring campaign so far has been the giant flare of HST-1 \cite{Har06},
which reached its maximum intensity in 2005 (Fig.~1) when the TeV emission was
detected in flaring state for the first time, suggesting HST-1 as the possible
origin of the VHE emission \cite{Che07}.  Simultaneously, a huge optical flare
was detected by the Hubble Space telescope \cite{HST_M87}.  Additional
observations were taken in 2007 on shorter intervals to investigate short-time
variability and possible correlation with VHE emission, which was unfortunately
in a quiet state at that time.

\begin{figure}
\includegraphics[width=0.49\textwidth]{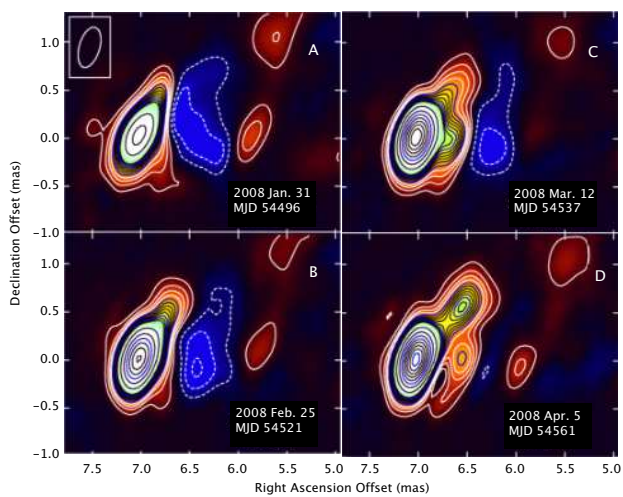}
\caption{
VLBA images of M 87 at 43 GHz. 
(A)-(D): Sequence of difference images for observations during the period of the radio 
flare.  These images have had an average image subtracted in order to show the effects of the flare.  
The average was made from eleven observations in 2007, outside the period affected by the flare.
The contours are linear with 10 (white) at intervals
of $7 \, \rm{mJy}$ per beam followed by the rest (black) at intervals of
$70 \, \rm{mJy}$ per beam; negative contours are indicated by dashed lines. 
The sequence shows the significant rise in the core flux density and the appearance of 
enhanced emission along the inner jet. 
\label{fig:radio}}
\end{figure}

\subsection{Radio: VLBA}
 Throughout 2007, M\,87 was observed at 43\,GHz with the VLBA on a regular
basis roughly every three weeks \cite{Supp_VLBA_MovieM87}.  In January 2008,
the campaign was intensified to one observation every 5 days.  The resolution
of the observations is rather high with 0.21$\times$0.43 milliarcseconds or
30$\times$60 Schwarzschild diameter of M\,87.  The aim of this `movie project'
was to study morphological changes of the plasma jet with time. Preliminary
analysis of the first 7 months showed a fast evolving structure, somewhat
reminiscent of a smoke plume, with apparent velocities of about twice the speed
of light. These motions were faster than expected so the movie project was
extended from January to April 2008 with a sampling interval of 5 days. 

\subsection{Results}

In January 2008, the VHE $\gamma$-ray flux was measured at a slightly higher
level than in 2007.  MAGIC detected a strong flaring activity in February 2008
\cite{Alb08,Err09}, which led to immediate intensified observations by all
three VHE experiments.  VERITAS detected another flare about one week after the
MAGIC trigger. The joint 2008 VHE $\gamma$-ray light curve clearly confirms the
short-term variability reported by H.E.S.S. in 2005 \cite{Aha06}. During the
2008 VHE flaring activity, MAGIC observed flux variability above $350 \,
\rm{GeV}$ on time scales as short as 1~day (at a significance level of
$5.6$~standard deviations). At lower energies ($150 \, \rm{GeV}$ to $350 \,
\rm{GeV}$) the emission was found to be compatible with a constant level
\cite{Alb08,Err09}.  From March to May, M\,87 was back in a quiet state.

In 2008, the X-ray and VHE $\gamma$-ray data suggested a different picture
compared to the 2005 flare (Fig.~\ref{fig:LC_All}): HST-1 was in a low state,
with the flux being comparable with the X-ray flux from the core. The core,
however, showed an increased X-ray flux state in February 2008, reaching a
highest flux ever measured with Chandra just few days after the VHE flaring
activity.

VLBA measured a continuously increasing radio flux from the region of the
nucleus ($r$=1.2 mas) during the 2008 campaign, whereas in 2007 the flux was
found to be rather stable.  Individual snapshots of the inner region of the jet
are shown in Fig.~\ref{fig:radio}.  The observed radio flux densities reached
at the end of the 2008 observations, roughly 2 months after the VHE flare
occurred, are larger than seen in any previous VLBI observations of M\,87 at
this frequency, including during the preceding 12 months of intensive
monitoring, in 6 observations in 2006 and in individual observations in 1999,
2000, 2001, 2002, and 2004 \cite{Supp_VLBA_M87}.  This suggests that radio
flares of the observed magnitude are uncommon.  The correlation study and the
implication of the 2008 results on the VHE emission models are presented in
detail in \cite{Acc09}.

\section{Conclusion and Outlook}
The cooperation between H.E.S.S., MAGIC and VERITAS in 2008 allowed for an
optimized observation strategy, which resulted in the detection and detailed
measurement of a VHE $\gamma$-ray outburst from M\,87.
Simultaneous Chandra observations, found HST-1, the innermost knot in the jet,
in a low state, while the nucleus showed increased X-ray activity. This is in
contrast to the 2005 VHE $\gamma$-ray flare, where HST-1 was in an extreme high
state.  The radio activity in 2007--2008, resolving the inner region of M87
down to some 10s Schwarzschild radii, allows for speculation about the origin
of the VHE $\gamma$-ray emission. 
A model suggesting HST-1 to be the origin of the $\gamma$-ray emission seems
less likely in the light of the 2008 result.  Due to its proximity and the
viewing angle of the jet, M\,87 is a unique laboratory for studying the
connection between jet physics and the measured VHE $\gamma$-ray emission. To
finally unravel the location/mechanism of the VHE $\gamma$-ray emission, future
coordinated VHE observations (such as the 2008 campaign) are essential,
together with simultaneous observations in the radio to X-ray (Chandra)
regimes.  Future joint observations could also lead to the detection of VHE
$\gamma$-ray intra-night variability, which would futher constrain the size of
the emission region.


\section*{Acknowledgments}
{\it H.E.S.S.:} The support of the Namibian authorities and of the
University of Namibia in facilitating the construction and operation of
H.E.S.S. is gratefully acknowledged, as is the support by the German
Ministry for Education and Research (BMBF), the Max Planck Society, the
French Ministry for Research, the CNRS-IN2P3 and the Astroparticle
Interdisciplinary Programme of the CNRS, the U.K.  Science and
Technology Facilities Council (STFC), the IPNP of the Charles
University, the Polish Ministry of Science and Higher Education, the
South African Department of Science and Technology and National Research
Foundation, and by the University of Namibia. We appreciate the
excellent work of the technical support staff in Berlin, Durham,
Hamburg, Heidelberg, Palaiseau, Paris, Saclay, and in Namibia in the
construction and operation of the equipment. \\
{\it MAGIC:} MAGIC acknowledges the excellent working conditions a the
Instituto de Astrofisica de Canarias' Observatorio del Roque de los Muchachos
in La Palma. The support of the German BMBF and MPG, the Italian INFN and
Spanish MCINN is gratefully acknowledged. This work was also supported by ETH
Research Grant TH 34/043, by the Polish MNiSzW Grant N~N203~390834, and by the
YIP of the Helmholtz Gemeinschaft. \\
{\it VERITAS:} 
This research is supported by grants from the US Department of Energy, the US
National Science Foundation, and the Smithsonian Institution, by NSERC in
Canada, by Science Foundation Ireland, and by STFC in the UK. We acknowledge
the excellent work of the technical support staff at the FLWO and the
collaborating institutions in the construction and operation of the instrument. \\
{\it VLBA team:} The Very Long Baseline Array is operated by The National Radio
Astronomy Observatory, which is a facility of the National Science Foundation
operated under cooperative agreement by Associated Universities, Inc.

\end{document}